\documentclass{article}

\usepackage{arxiv}

\usepackage[utf8]{inputenc} % allow utf-8 input
\usepackage[T1]{fontenc}    % use 8-bit T1 fonts
\usepackage{hyperref}       % hyperlinks
\usepackage{url}            % simple URL typesetting
\usepackage{booktabs}       % professional-quality tables
\usepackage{amsfonts}       % blackboard math symbols
\usepackage{nicefrac}       % compact symbols for 1/2, etc.
\usepackage{microtype}      % microtypography
\usepackage{lipsum}
\usepackage{graphicx}
\usepackage{amsbsy, amstext, amssymb, amsthm, amsmath,bm,bbm,wasysym}
\usepackage{comment}
\usepackage{accents}

\usepackage{natbib}

\graphicspath{ {./figs/} }

\newcommand{\Abf}{{\bm A}}

\newcommand{\Xbf}{{\bm X}}

\newcommand{\hbf}{{\bm h}}

\newcommand{\xbf}{{\bm x}}

\newcommand{\greekbold}[1]{\mbox{\boldsymbol{$#1$}}}
\newcommand{\alphabf}{\greekbold{\alpha}}
\newcommand{\betabf}{\greekbold{\beta}}

\newcommand{\thetabf}{\greekbold{\theta}}

\title{Addressing errors in multiple variables using generalized raking and cumulative probability models}

\author{
 Eric S. Kawaguchi\\
  Department of Population and Public Health Sciences \\
  University of Southern California\\
  \texttt{ekawaguc@usc.edu} \\
  \And
 Chun Li \\
  Department of Population and Public Health Sciences \\
  University of Southern California\\
    \And
   Frank E. Harrell Jr.\\
  Department of Biostatistics \\
  Vanderbilt University \\
    \And
   Pamela A. Shaw\\
  Biostatistics Division \\
  Kaiser Permanente Washington Health Research Institute  \\
    \And
   Thomas Lumley\\
  Department of Statistics\\
  University of Auckland \\
  \And
   Bryan E. Shepherd\\
  Department of Biostatistics \\
  Vanderbilt University \\
}

\begin{document}
\maketitle

\begin{abstract}

Routinely collected data, such as electronic health record (EHR) data, are frequently used for biomedical research, but these data are prone to errors, which can bias study findings. Validating data in subsamples of records can reduce bias, and the efficiency of estimates can be improved by incorporating in analyses both the error-prone data available on the entire cohort and the validated data available on the subsample. One approach to incorporate both data sources is with generalized raking, which calibrates validation sampling weights using error-prone data from the entire cohort. Motivated by an EHR study of maternal weight gain during pregnancy with a validation subsample, we develop and illustrate generalized raking techniques for cumulative probability models (CPMs). CPMs are robust, rank-based and semiparametric models for continuous, ordinal, or mixed type outcome data. We develop efficient generalized raking estimators for CPMs, evaluate their performance relative to competing methods, and demonstrate the utility and strengths of generalized raking with CPMs in a study that examines factors associated with weight gain during pregnancy.

\end{abstract}

% keywords can be removed

\section{Introduction}

Routinely collected data, such as electronic health record (EHR) data, are increasingly being used for biomedical research. However, these data are prone to errors, which could result in misleading conclusions if used naively in medical studies \cite{floyd2012, weng2012}. The quality of EHR data can be greatly improved with careful review of medical charts and data validation. However, due to cost and time constraints, it is often not feasible to correct all records in the EHR. Instead, researchers may validate data on a sub-sample of records. The two levels of data collection are referred to as double-sampling or two-phase sampling in the statistical literature \cite{cochran77, sarndal2003model}. Phase-1 data are the EHR data and phase-2 data are the validation subsample. 

As an example, researchers are interested in factors associated with maternal weight gain during pregnancy. Data from 10,132 women who delivered singletons at Vanderbilt University Medical Center were extracted from the EHR. Due to data quality concerns, a research nurse reviewed the charts for a probabilistic sample of 726 (7.2\%) of these women. Errors were found in nearly all variables, including the outcome and covariates of interest, for a substantial percentage of records (Table \ref{tab:errors}). We would like to perform an analysis that incorporates the validated data to reduce the bias of our estimates, but also incorporates the much larger EHR dataset to improve statistical efficiency. Classical methods for addressing measurement error, e.g., regression calibration, moment-based methods, or SIMEX, are unable to handle our setting with errors in many variables \cite{carroll06}. Multiple imputation, where missing validated data are multiply imputed based on models fit to the validation data, can address errors across many variables \cite{cole06, giganti20}. However, multiple imputation estimators can be biased if the imputation models are incorrectly specified, which is likely in practice \cite{han21}.

Generalized raking (GR) is a robust and efficient method for combining EHR and validated data in the presence of errors across multiple variables \cite{oh20}. GR, also known as survey calibration, was originally developed in the survey sampling literature \cite{deville92, sarndal2003model} but has more recently been recognized in the biostatistics literature as a simple approach for obtaining augmented inverse probability weighted (AIPW) estimators \cite{lumley11}.  In short, GR calibrates phase-2 sampling weights using auxiliary information available in the phase-1 sample. The information from the phase-1 sample is brought in to improve efficiency of estimators while making no additional assumptions beyond those made by classical IPW estimators. The optimal auxiliary variables for calibrating weights are the expected value of the influence functions for the parameters of interest given the phase-1 (error prone) data \cite{breslow09}. %are the expected values of the influence functions for parameters of interest given phase-1 data \cite{breslow09}. 
These expectations are typically unknown, but they can be estimated using the influence functions from a model fit to the naive, phase-1 data. GR has been seen to be a robust, efficient, and flexible approach with minimal assumptions that can be used for a wide number of statistical models in two-phase settings with errors in a large number of variables \cite{lumley11, oh20, han21, shepherd23}.

\begin{figure}[h]
\centering
\includegraphics[scale=1]{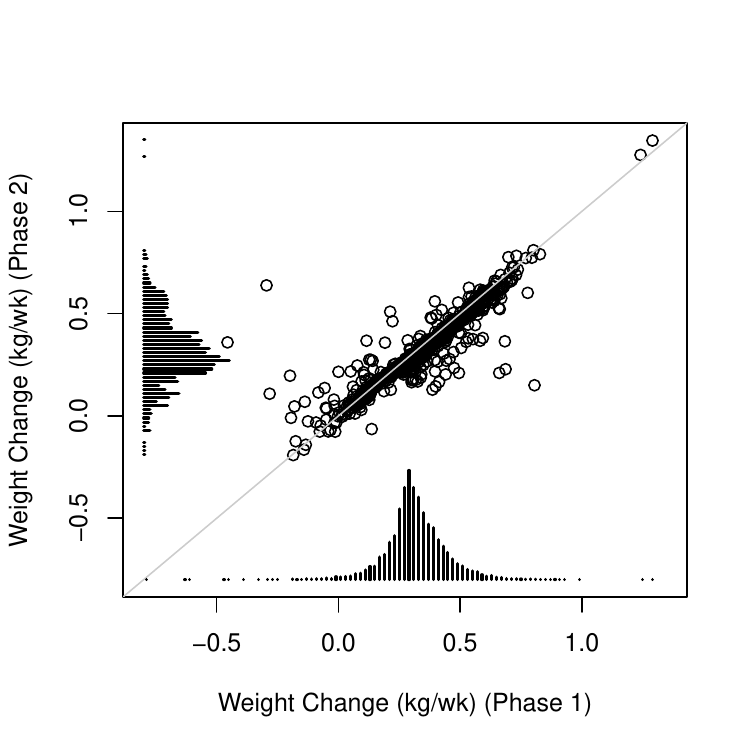}
\caption{Estimated weight change during pregnancy based on the error-prone electronic health record data (phase-1 cohort, $N=10,132$) and among the validation subset (phase-2 cohort, $n=726$). The histograms show the marginal distributions, and the scatterplot shows the relationship between phase-1 and phase-2 estimates for those in the validation subset.}
\label{fig:weight-change-correlation}
\end{figure}

With outliers or skewed continuous outcomes, rank-based analyses are often desirable. For example, maternal weight gain during pregnancy, quantified in our study as the estimated weight change per week (i.e., the estimated weight at delivery minus the estimated average weight at conception divided by the estimated pregnancy length), is slightly right skewed with modest outliers in both the phase-1 and phase-2 data (Figure \ref{fig:weight-change-correlation}). A robust, rank-based regression method is the cumulative probability model (CPM) \cite{liu2017}. The CPM assumes a linear model on a latent variable arising from a monotonic transformation of the response variable. The transformation is not specified and is estimated from the data. With continuous response data, the CPM is a straightforward method for fitting semiparametric linear transformation models \cite{zeng2007}. The CPM is equivalent to assuming that the continuous response variable is ordinal, with each unique outcome corresponding to an ordinal level, and fitting an ordinal regression model (e.g., the proportional odds model \cite{walker1967,mccullagh1980,agresti10}) to the continuous response data. Recognizing that large sections of the Hessian matrix are zero, %Cholesky decomposition can be used to speed up computation, and 
CPMs can be fit in seconds to data with tens of thousands of unique response variables using ordinal regression models \cite{liu2017}. From a fitted CPM, conditional odds ratios, expectations, quantiles, and exceedance probabilities are easily derived, yielding robust and interpretable results.

Our goal is to investigate factors associated with maternal weight gain during pregnancy using a CPM that applies generalized raking techniques to incorporate validation data to account for errors across multiple variables in the EHR. This manuscript combines two novel and robust techniques -- generalized raking and CPMs -- to address an important scientific question. Although sampling weights have been applied to ordinal regression models in settings with a relatively small number of categories \cite{natarajan2012,liu2016,mitani24}, we are unaware of their application to CPMs fit to continuous data. In addition, calibration of survey weights using estimated influence functions from CPMs as auxiliary variables has never been performed, and its application with continuous responses is non-trivial. This novel approach for addressing errors across multiple variables results in efficient and interpretable estimates.

The remainder of this manuscript is organized as follows: In Section \ref{sec:motivation} we provide a more detailed description of our motivating example. In Section \ref{sec:methods} we describe our method, including reviews of CPMs and generalized raking, and then how to combine the two. We discuss estimation, computation, and inference. In Section \ref{sec:rda} we illustrate the use of our methods to investigate factors associated with maternal weight gain during pregnancy. In Section \ref{sec:sim} we present simulations that illustrate the excellent performance of our method. Finally, in Section \ref{sec:discussion} we discuss our findings and suggest directions for future research. R code for simulations and real-data analysis, as well as a synthetic maternal weight gain dataset, are available in the Supplemental Material.

\begin{table}[h]
\centering
\resizebox{\textwidth}{!}{%
\begin{tabular}[t]{lllll}
\toprule
Variable & Phase 1 ($N=10,132$) & Phase 2 ($n=726$) & Proportion in error & Discrepancy\\
\midrule

Weight gain (kg/week) &  0.31 (0.26, 0.38) & 0.31 (0.22, 0.41) & 1.00 & 0.930; -0.02 (range -0.82, 0.66) \\

Body mass index (kg/m$^2$) &  25.9 (22.6, 30.5) & 27.8 (23.7, 33.1) & 1.00 & 0.985; -0.13 (range -8.6, 6.8) \\ 

Age at delivery (years) &  28.0 (23.4, 32.3) &  27.5 (23.0, 31.8) & 0 & --  \\ 
           
Race    & & & 0.06 \\

\hspace{.2in} Asian &  702 (0.07) &  40 (0.04) &  & PPV=0.871, NPV=0.997  \\           

\hspace{.2in} Black &  2341 (0.23) &  288 (0.30) &  & PPV=0.990, NPV=0.992  \\           

\hspace{.2in} White &  6255 (0.62) &  544 (0.56) &  & PPV=0.951, NPV=0.952  \\           

\hspace{.2in} Other/Unknown &  834 (0.08) &  93 (0.10) &  & PPV=0.746, NPV=0.965  \\          

Hispanic Ethnicity &  1520 (0.15) & 113 (0.16) & 0.01 & PPV=0.941, NPV=0.998\\ 

Tobacco Use during Pregnancy &  641 (0.06) &  101 (0.14) & 0.11 & PPV=0.705, NPV=0.897  \\ 

Depression &  899 (0.09) &  69 (0.10) & 0.13 & PPV=0.321, NPV=0.933  \\ 
           
Private Insurance &  5472 (0.54) &  231 (0.32) & 0.25 & PPV=0.572, NPV=0.934 \\ 

Length of Pregnancy (weeks)  &  -- &  39.3 (38.3, 40.3) & -- & --  \\ 
  
\bottomrule
\end{tabular}%
}
\caption{Study variables and their error rates. Columns summarizing characteristics of the phase-1 and phase-2 data include counts (proportions) for binary variables and medians (25th, 75th percentiles) for continuous variables. The discrepancy column shows the positive predictive value (PPV) and negative predictive value (NPV) for binary variables and for continuous variables shows the correlation; median difference and range of difference (phase-1 value minus phase-2 value). Length of pregnancy was only collected in phase-2 data.}
\label{tab:errors}
\end{table}

%\begin{figure}[h]
%\includegraphics[scale=0.75]{weight-change-boxplot.pdf}
%\caption{Boxplot of estimated weight change during pregnancy based on the error-prone electronic health record data (phase-1 cohort, $N=10,132$) and among the validation subset (phase-2 cohort, $n=726$).}
%\label{fig:weight-change-boxplot}
%\end{figure}

\section{Motivating Example: Maternal weight gain during pregnancy}
\label{sec:motivation}
Although some maternal weight gain during pregnancy is, of course, natural and desirable, excessive weight gain has been linked to childhood obesity and poor maternal health, including obesity and cardiometabolic disease \cite{goldstein17}. Researchers are interested in investigating factors that are potentially related to maternal weight gain during pregnancy. Risk factors that were a priori specified as of potential interest included body mass index prior to pregnancy, age, race, ethnicity, tobacco use during pregnancy, depression, insurance status, and the length of the pregnancy. 

An earlier study investigated the association between maternal weight gain during pregnancy and childhood obesity among mother-child pairs delivering at the Vanderbilt University Medical Center \cite{shepherd23,sneed24}. EHR data were extracted. Because of the error-prone nature of the EHR data, chart reviews on a subsample of records were performed by a research nurse. Charts were selected for validation in a multi-wave, probabilistic manner designed to maximize the precision of the adjusted association between maternal weight gain and childhood obesity. To this end, 33 strata were created based on categorizations of phase-1 variables including maternal weight gain, childhood obesity, and child follow-up time. Researchers sampled without replacement records from strata for chart review. Details on strata and sampling are provided elsewhere \cite{shepherd23}. A secondary study created and sampled from new strata to investigate links between maternal weight gain and childhood asthma. However, due to tangential complexities that arise from combining data across separate sampling frames \cite{metcalf09}, in this study we only consider those sampled for data validation as part of the obesity sampling.

A total of $N=10,132$ women giving birth to singletons were included in the study and are referred to as the phase-1 cohort. Among these women, charts were reviewed for $n=726$, referred to as the phase-2 cohort. In both phase-1 and phase-2 data, maternal weight trajectories were estimated for each woman using a functional principal components analysis; the estimated maternal weights at conception and delivery were then extracted from these models \cite{ramsay07,yao05,shepherd23}. A summary of phase-1 and phase-2 data, as well as error rates based on the chart review, are shown in Table \ref{tab:errors}. Figure \ref{fig:weight-change-correlation} shows the distribution of phase-1 and phase-2 weight gains during pregnancy. All study variables extracted from the EHR, except age at delivery, were discovered from the chart reviews to have errors. The extent and magnitude of errors varied substantially across variables. Weight gain during pregnancy and BMI at start of pregnancy were discovered to be incorrect in 100\% of validated phase-1 records, although the correlation with validated values was 0.930 and 0.985, respectively. The discrepancies were primarily because length of pregnancy, which was used in the derivation of weight change during pregnancy and BMI at conception, was unavailable in the phase-1 data and was uniformly assumed to be 39 weeks. Pregnancy length was corrected in the phase-2 data after chart review.  Insurance status was misclassified in 25\% of validated records, and depression (13\%) and tobacco use during pregnancy (11\%) also had non-trivial error rates.

Errors of this frequency and magnitude could lead to biased estimation if we rely only on phase-1 data for analyses. However, there is rich (although error-prone) information in the EHR that we do not want to discard from a large number of women; hence, we want to perform analyses using more than the relatively small subset of validated records. Generalized raking will allow us to incorporate both phase-1 and phase-2 data into our analyses with minimal assumptions. In addition, we are interested in robust, rank-based estimates of association, yet also interpretable quantities such as medians (or other quantiles) and exceedance probabilities conditional on covariates. CPMs will allow us to obtain such estimates. We have now set the stage for introducing our proposed statistical methods.

\section{Methods}
\label{sec:methods}

\subsection{Review of CPMs for continuous outcomes}
\label{sec:cpm_review}

The CPM extends the widely-used ``cumulative link" model for ordered multinomial data to continuous outcomes.  Let $Y$ be a continuous outcome and $\Xbf= (X_{1}, \ldots, X_{p})^T$ be a vector of $p$ covariates. 
The CPM models the cumulative distribution function of $Y$ conditional on $\Xbf$ through
\begin{align}
\label{eq:cpm}
    G[\Pr(Y \leq y| \Xbf= \xbf)] = \alpha(y) - \betabf^T\xbf,
\end{align}
where $G(\cdot)$ is a link function mapping values from $(0, 1)$ to $\mathbb{R}$ and $\alpha(\cdot)$ is an unspecified monotonically increasing intercept function. 

Let $(Y_i,\Xbf_i)$ be independent and identically distributed draws of $(Y,\Xbf)$ for $i = 1, \ldots, N$. Without loss of generality, assume each observed value of $Y_i$, denoted $y_i$, is ordered such that $y_1 \leq y_2 \leq \ldots \leq y_N$. 
Given the observed outcomes $\{y_i\}_{i=1}^N$, the CPM can be expressed as 
\begin{align}
\label{eq:cpm2}
    G[\Pr(Y \leq y_i| \Xbf= \xbf)] = \alpha_i - \betabf^T\xbf,
\end{align}
with parameters $(\alpha_1, \alpha_2, \ldots, \alpha_N, \betabf)$, where $\alpha_i = \alpha(y_i)$. Since $\alpha(\cdot)$ is increasing, we have the constraint $\alpha_1 \leq \alpha_2 \leq \ldots \leq \alpha_N$ and $\alpha_i = \alpha_j$ whenever $y_i = y_j$.

The regression coefficients $\betabf$ correspond to (adjusted) stochastic changes in the distribution of $Y$. Specific choices of the link function $G$ can lead to interpretable measures of associations. For example, when $G$ is the logit link, (\ref{eq:cpm}) simplifies to the standard ``proportional odds model" and $\betabf$ are log-odds ratios. The CPM is equivalent to a semiparametric linear transformation model: after an unspecified transformation, $\alpha(\cdot)$ which is estimated non-parametrically with a step function, the transformed outcome follows a linear model with mean $\betabf^T\xbf$ and errors following the distribution specified by the chosen link function (e.g., if the logit link function is used, the errors are assumed to follow a standard logistic distribution) \cite{zeng2007}. It is apparent from (\ref{eq:cpm}) that unlike linear regression or quantile regression, which only focus on \textit{one} aspect of the conditional distribution (i.e., the conditional mean or quantile), the CPM models the \textit{entire} conditional distribution. Furthermore, the CPM enforces monotonicity of the estimated conditional quantile function, whereas estimating quantile regression at multiple quantiles independently does not guarantee this property.

Liu et al. \cite{liu2017} showed that the nonparametric likelihood function for the CPM can be written as
\begin{align}
    \label{eq:alik}
    L(\alphabf, \betabf) & = \prod_{i=1}^N L_i(\alphabf, \betabf) \notag \\
    & = \prod_{i:y_i=a_1} [G^{-1}(\alpha_1 - \betabf^T\xbf_i)] \times \prod_{i:y_i=a_J} [1 - G^{-1}(\alpha_{J-1} - \betabf^T\xbf_i)] \notag \\
    & \times \prod_{i=2}^{J-1} \prod_{i:y_i=a_j} [G^{-1}(\alpha_j - \betabf^T\xbf_i) - G^{-1}(\alpha_{j-1} - \betabf^T\xbf_i)],
\end{align}
where $\{a_1,\ldots,a_J\}$ represent the $J$ distinct observed outcome values for $y_i$, $\alpha_j = \alpha(a_j)$ for $j = 1, \ldots, J$, and $\alpha_0 = -\infty$ and $\alpha_J = \infty$, respectively. In the absence of ties (i.e., $J=N$), the likelihood simplifies to 
\begin{align*}
    L(\alphabf, \betabf) 
    & = [G^{-1}(\alpha_1 - \betabf^T\xbf_1)] \times [1 - G^{-1}(\alpha_{N-1} - \betabf^T\xbf_N)] \times \prod_{i=2}^{N-1}[G^{-1}(\alpha_i - \betabf^T\xbf_i) - G^{-1}(\alpha_{i-1} - \betabf^T\xbf_i)].
\end{align*}
The likelihood (\ref{eq:alik}) has the same structure as the multinomial likelihood of the CPM for discrete ordered outcomes where each distinct value of $Y$ is its own category \cite{agresti10}. This equivalence allows us to obtain the NPMLE for $(\alphabf, \betabf)$ by treating $Y$ as a discrete ordered variable and fitting the discrete CPM using readily-available software for ordered multinomial regression. Recognizing that large sections of the score and Hessian matrices are zero \cite{liu2017}, Cholesky decomposition can be used to speed up computation, and CPMs can be fit in seconds to data with tens of thousands of unique response variables using the \texttt{orm} function in the R package \texttt{rms} \cite{rms}. Under mild regularity assumptions and assuming a bounded range of the continuous response variable, CPM estimates are consistent, asymptotically normal, and their variance can be estimated using the inverse of the Hessian matrix \cite{li2023}. Extensive simulations have demonstrated good finite sample performance of CPMs, the robustness of CPMs to moderate link function misspecification, and the utility of CPMs for modeling response data that are skewed, contain outliers, or are a mixture of continuous and discrete variables (e.g., outcomes with detection limits) \cite{liu2017, tian2020, li2023, tian2024addressing}.

\subsection{CPMs with generalized raking under two-phase sampling}
\label{sec:twophase_cpm}

In our setting, the EHR does not contain $(Y_i,\Xbf_i)$, but error-prone versions of these variables, denoted as $(Y_i^*,\Xbf_i^*)$. Let $R_i$ be the indicator that subject $i$ was selected for validation. If $R_i=1$, then $(Y_i,\Xbf_i)$ are known, otherwise $(Y_i,\Xbf_i)$ are missing. In the maternal weight gain study, we selected records for chart review using stratified random sampling, where strata were defined based on $(Y_i^*,\Xbf_i^*)$. Thus $(Y_i,\Xbf_i)$ are missing at random (MAR), i.e., $\Pr(R_i=1|Y_i^*,\Xbf_i^*, Y_i, \Xbf_i) = \Pr(R_i=1|Y_i^*,\Xbf_i^*)$. Let $\pi_i \equiv \Pr(R_i=1|Y_i^*,\Xbf_i^*)$ and $n \equiv \sum_{i=1}^N R_i.$ 

Consider the CPM defined by (\ref{eq:cpm2}). For notational convenience, let $\thetabf = (\alphabf, \betabf)$. An inverse-probability weighted (IPW) estimator of $\thetabf$ is obtained by maximizing the weighted log-likelihood (\ref{eq:alik}) in the validation subset. Specifically, 
IPW estimates, denoted $\hat{\thetabf}_w = (\hat{\alphabf}_w, \hat{\betabf}_w)$, are obtained by solving
\begin{align}
  \label{eq:ee}
  \sum_{i=1}^N w_i R_i U_i(\thetabf)=0,
\end{align}
where $w_i=\pi_i^{-1}$ is the sampling weight, and $U_i(\thetabf) = \frac{\partial}{\partial \thetabf} \log(L_i(\thetabf))$ is the score equation for the $i$th individual. Cholesky decomposition algorithms can be applied to maximize the weighted log-likelihood just as they are applied to maximize the unweighted likelihood (\ref{eq:alik}), thus ensuring fast computation. Under standard assumptions including MAR and positivity (i.e., $\pi_i>0$ for all $i$), $\hat{\thetabf}_w$ is consistent for $\thetabf.$ However, the variance of $\hat{\thetabf}_w$ may be large.

Generalized raking (GR) is a method that improves the efficiency of IPW estimators by calibrating the inverse-probability weights using auxiliary variables, $\Abf$, available in the phase-1 cohort. The calibrated weights are defined as $w_i=g_i\pi_i^{-1}$, 
where $g_i$ are obtained by minimizing
\begin{align}
\label{eq:gr}
    \sum_{i=1}^N R_i d(\pi_i^{-1},g_i \pi_i^{-1}) \hspace{.3in} \mbox{subject to  } \sum_{i=1}^N g_i\pi_i^{-1}R_i\Abf_i = \sum_{i=1}^N \Abf_i,
\end{align}
and $d(\cdot)$ is a loss function $d(a,b)=a\log(a/b)+(b-a)$. In this way, GR tweaks the inverse-probability weights, keeping them as close as possible to their original values while making the weighted total of $\Abf$ in the phase-2 subsample exactly equal to the known total of $\Abf$ in the phase-1 sample. Although $\Abf$ can be arbitrary, to increase efficiency, the choice of $\Abf$ should be based on the target parameters.
%it should be positively correlated with the influence functions for the parameters estimates based on the true data.

In our context, the optimal auxiliary variable for estimating $\thetabf$ in the CPM using GR is $\Abf_i = E[\hbf(\thetabf; Y_i, \Xbf_i)|Y_i^*,\Xbf_i^*]$ for influence function $\hbf(\thetabf; Y_i, \Xbf_i)$ \cite{breslow09, oh20}. The rationale for this choice is as follows. Suppose $(Y_i, \Xbf_i)$ is available for all subjects in our phase-1 cohort.
Then $\hat{\thetabf}$, obtained by maximizing (\ref{eq:alik}), can be expressed as:
\begin{align}
    \sqrt{N}(\hat{\thetabf} - \thetabf) = \sum_{i=1}^N \hbf(\thetabf; Y_i, \Xbf_i) + o_p\left(N^{-1/2}\right),
\end{align}
where $\hbf(\thetabf; Y_i, \Xbf_i) = -\mathcal{I}(\thetabf)^{-1}U_i(\thetabf),$ 
and $\mathcal{I}(\thetabf)$ is the Fisher information of (\ref{eq:alik}). In our setting, we do not know $(Y_i,\Xbf_i)$ for all phase-1 records
%so we cannot calibrate with the optimal $E[\hbf(\thetabf; Y_i, \Xbf_i)|Y_i^*,\Xbf_i^*]$
but we do know $(Y_i^*,\Xbf_i^*)$, so $E[\hbf(\thetabf; Y_i, \Xbf_i)|Y_i^*,\Xbf_i^*]$ is our best proxy of $\hbf(\thetabf; Y_i, X_i)$ based on the available data and therefore it is the optimal auxiliary variable for estimating $\thetabf$. Since this quantity is based on unknown parameters, we approximate it with $\hbf(\hat{\thetabf}^*; Y_i^*, \Xbf_i^*)$, where $\hat{\thetabf}^*$ is estimated by maximizing (\ref{eq:alik}) based on $(Y_i^*, \Xbf_i^*)$. We then optimize (\ref{eq:gr}) using $\hbf(\hat{\thetabf}^*; Y_i^*, \Xbf_i^*)$ as the auxiliary variable to obtain $g_i$. Finally, our GR estimator, $\hat{\thetabf}_g$, is the solution to equation (\ref{eq:ee}) with $w_i=g_i \pi_i^{-1}$.  %In practice, the naive influence function is estimated by maximizing (\ref{eq:alik}) but using $(Y_i^*,\Xbf_i^*)$ instead of $(Y_i,\Xbf_i)$, and then estimating influence functions based on this fitted model.  

One can compute $\hbf(\hat{\thetabf}^*; Y_i^*, \Xbf_i^*)$, which we refer to as the `naive influence function' \cite{shepherd23}, as $-\mathcal{I}(\hat{\thetabf}^*)^{-1}U_i(\hat{\thetabf}^*)$, where $\mathcal{I}(\hat{\thetabf}^*)$ and $U_i(\hat{\thetabf}^*)$ are the information matrix and subject-specific score vector based on $\hat{\thetabf}^*$, respectively. Because $\hat{\thetabf}^*$ is high dimensional for continuous $Y$, computing $\hbf(\hat{\thetabf}^*; Y_i^*, \Xbf_i^*)$ will require $\mathcal{O}(N^3)$ operations and solving the calibration equation (\ref{eq:gr}) with $\Abf_i=\hbf(\hat{\thetabf}^*; Y_i^*, \Xbf_i^*)$  may not be feasible.  Therefore, to efficiently calculate the influence functions, we take advantage of the facts that estimation of $\betabf$ is of primary interest, that the influence function for $\thetabf$ can be decomposed as $\hbf(\thetabf; Y_i, \Xbf_i)=(\hbf_{\alphabf}(\thetabf; Y_i, \Xbf_i ), \hbf_{\betabf}(\thetabf; Y_i, \Xbf_i))$, and that $\mathcal{I}(\thetabf)$ and $U_i(\thetabf)$ have sparse structures that can be leveraged. Details are provided in the Appendix. In practice, we calibrate our weights using only the estimated $p$-dimensional naive influence function for $\betabf$, denoted as $\hbf_{\betabf}(\hat{\thetabf}^*; Y_i^*, \Xbf_i^*)$.

We also use influence functions to quantify uncertainty of our weighted estimators $\hat{\thetabf}_w$ and $\hat{\thetabf}_g$.  These estimators are asymptotically normal and can be written as the weighted sum of the influence functions based on their weighted likelihood\cite{chen20}. Specifically, the GR estimator can be written as:
\begin{align*}
    \sqrt{n}(\hat{\thetabf}_g - \thetabf) = \sum_{i=1}^N w_i R_i \hbf^{(w)}(\thetabf; Y_i, \Xbf_i) + o_p\left(n^{-1/2}\right).
\end{align*} 
where $w_i=g_i \pi_i^{-1}$ and the weighted influence function is 
\begin{align}
    \label{eq:influence}
    \hbf^{(w)}(\thetabf; Y_i, \Xbf_i) = -\hat{\mathcal{I}}^{(w)}(\thetabf)^{-1}U^{(w)}_i(\thetabf),
\end{align}
where $\hat{\mathcal{I}}^{(w)}(\thetabf)$ is the weighted estimate of the population Fisher information and $U^{(w)}_i(\thetabf)$ is the score of the weighted likelihood for the $i$th individual. Therefore, the variance of $\hat \thetabf_g$ can be estimated via Taylor-series linearization as the sum of two components: 1) the inverse of the estimated full-cohort Fisher information defined as 
\begin{align*}
    I_1 = \left(\sum_{i=1}^N w_i R_i \hat{\mathcal{I}}^{(w)}_i(\thetabf) \right)^{-1}
\end{align*}

and 2) the estimated phase-two sampling variance
\begin{align*}
    I_2 =\sum_{i,j} \frac{R_i R_j w_i w_j}{\pi_{ij}} \hbf_i^{(w)}(\thetabf; Y_i, \Xbf_i) \hbf_j^{(w)}(\thetabf; Y_j, \Xbf_j)^T cov(R_i, R_j), 
\end{align*}
where $\pi_{ij} \equiv \Pr(R_i=1,R_j=1|Y_i^*,\Xbf_i^*,Y_j^*,\Xbf_j^*)$. Therefore,
\begin{align*}
    \widehat{\text{var}}(\hat \thetabf_g)= I_1 + I_2.
\end{align*}

\section{Application}
\label{sec:rda}

We applied our methods to investigate factors associated with weight gain during pregnancy. As mentioned earlier, of $N=10,132$ women in the phase-1 cohort, $n=726$ had a thorough chart review. Thirty-three strata were created based on phase-1 data, and charts were randomly selected for review within strata. All records that were selected for data validation were able to be validated. Therefore, validation data on the records that were not validated are missing at random (MAR). Sampling weights, $1/\pi_i$, were simply the inverse probability of being sampled based on the number of records in a stratum and the number of records sampled from that stratum.

To calculate naive influence functions for generalized raking, we fit a CPM model with a logit link function to the phase-1 data. Results from this model also serve as a comparator, showing us what we would have estimated had we not performed data validation. The CPM included phase-1 values of estimated weight change per week (outcome) regressed on mother's estimated BMI at the start of pregnancy, maternal age at delivery, maternal race (White [reference], Asian, Black, or Other/Unknown), Hispanic ethnicity (yes/no), tobacco use during pregnancy (yes/no), depression diagnosis (yes/no), and private insurance status (yes/no). From this fitted model, we extracted estimated influence functions for each covariate coefficient for each observation, i.e., $\hbf_{\betabf}(\hat{\thetabf}^*; Y_i^*, \Xbf_i^*)$. These naive influence functions were then used as auxiliary variables to calibration our inverse probability weights in the generalized raking procedure. Weights were calibrated as described in Section \ref{sec:twophase_cpm}. %such that $\sum_{i=1}^N g_i\pi_i^{-1}R_i \hbf_{\betabf}(\hat{\thetabf}^*; Y_i^*, \Xbf_i^*) = \sum_{i=1}^N \hbf_{\betabf}(\hat{\thetabf}^*; Y_i^*, \Xbf_i^*)$ while minimizing the loss $\sum_{i=1}^N R_i d(\pi_i^{-1},g_i \pi_i^{-1})$. 
In our analysis, this amounts to calibrating a vector of length nine, corresponding to the different covariates. We also included the strata in our calibration equations (again with a logit link function), such that the weighted sum of records in each stratum in the phase-2 sample equaled their weighted totals in the phase-1 sample. There were no issues with convergence, and we were thus able create calibrated weights, $g_i\pi_i^{-1}$. %that resulted in the weighted sum of all nine naive influence functions in the phase-2 sample being equal to their unweighted sum (zero, which is a property of influence functions) in the phase-1 sample. 
We finally fit a CPM, weighted by the calibrated weights, to the phase-2 data. In addition to the covariates listed above, our CPM included the estimated length of pregnancy, which was not available in the phase-1 data but was extracted from the EHR as part of the chart review process and thus was available in the phase-2 data. For comparison, we also computed IPW estimates for the CPM using the uncalibrated weights.

\begin{figure}[h]
\includegraphics[scale=1]{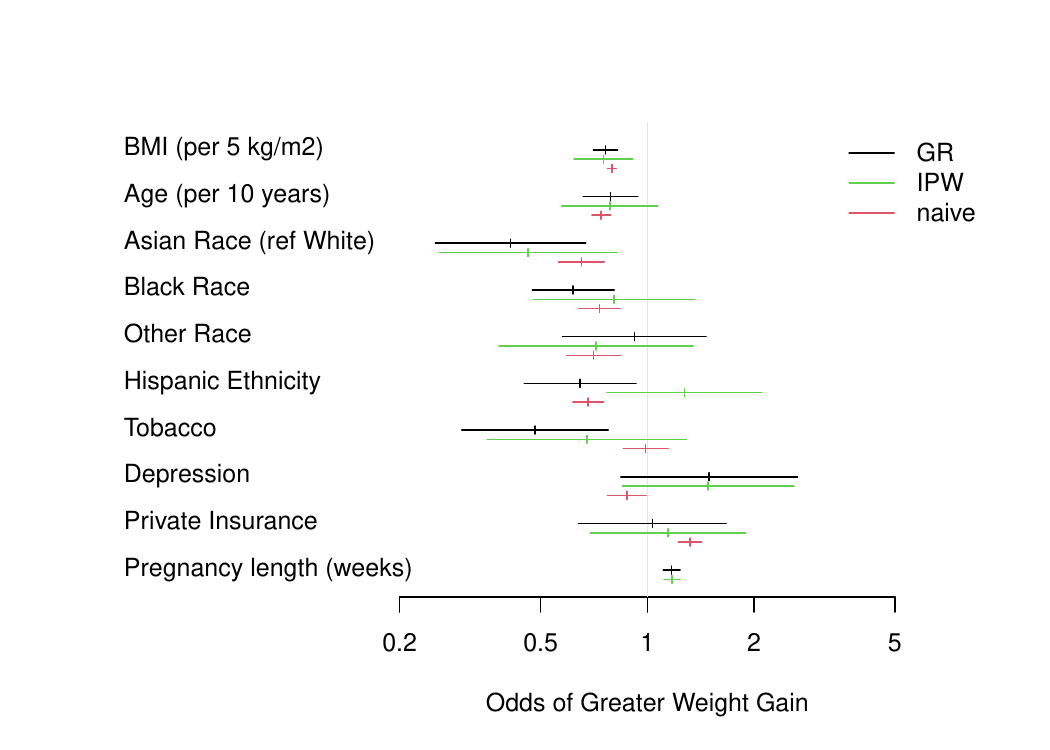}
\caption{Forest plot of odds ratios.}
\label{fig:odds-ratios}
\end{figure}

\begin{figure}[h]
\includegraphics[scale=1]{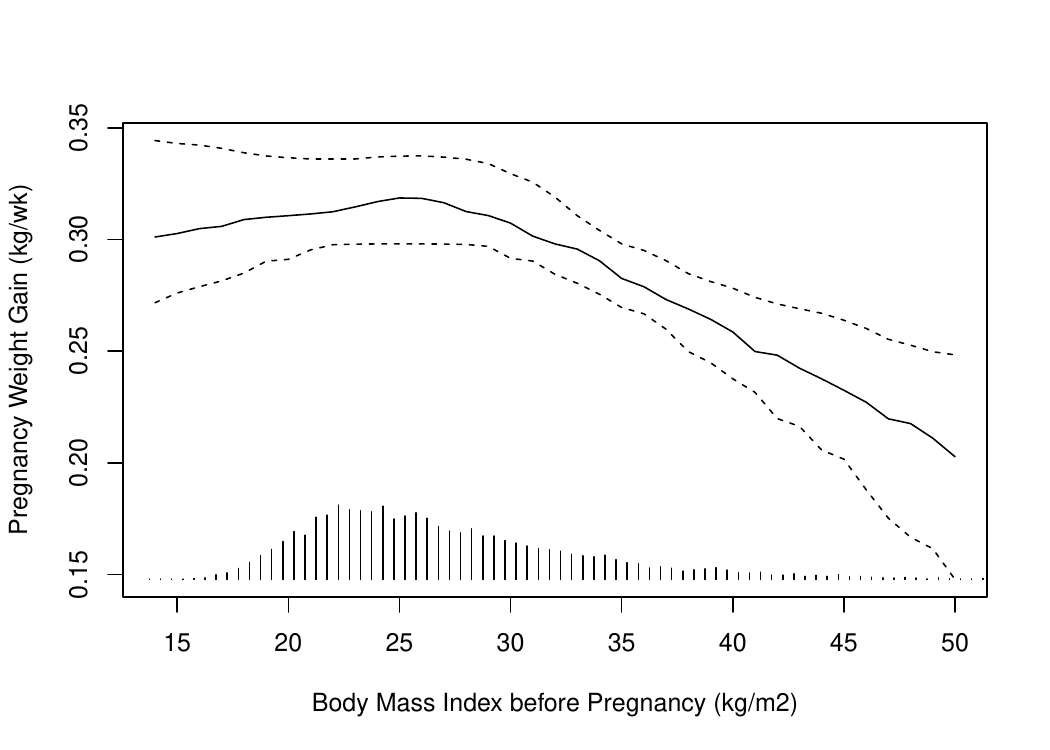}
\caption{Estimated median weight change (kg/wk) during pregnancy as a function of body mass index before pregnancy. 95\% confidence intervals for the median are included. A histogram of body mass index in the phase-1 cohort is shown at the bottom of the figure. All other covariates are set at their medians or modes.}
\label{fig:med-BMI}
\end{figure}

\begin{figure}[h]
\includegraphics[scale=1]{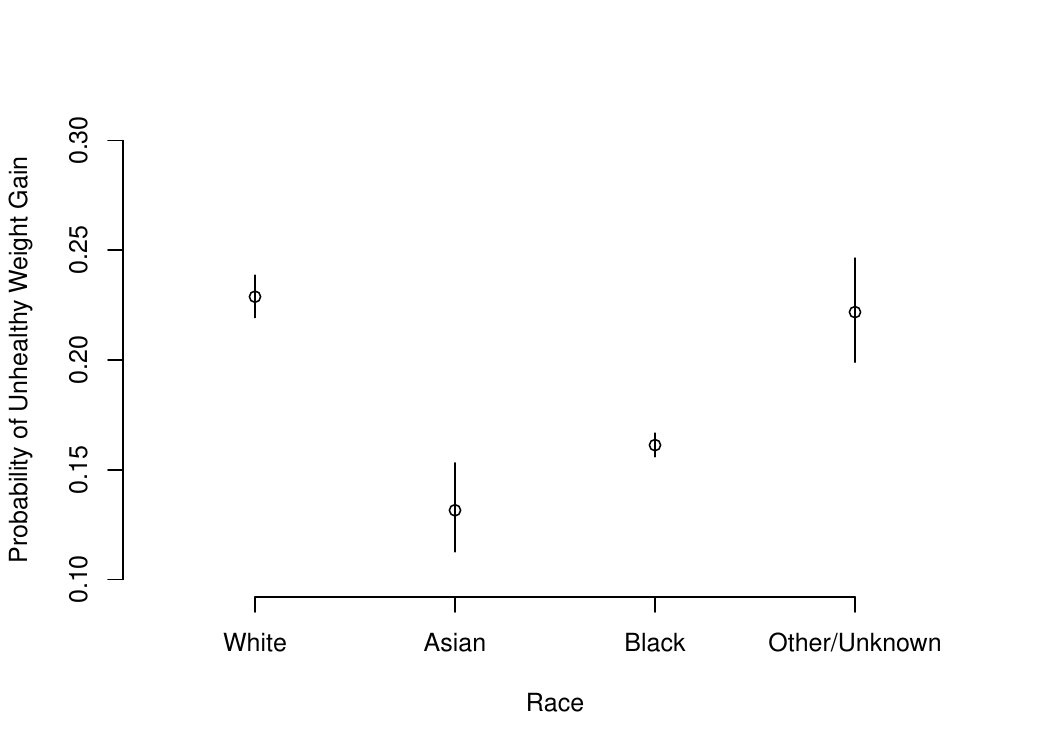}
\caption{Estimated probability of unhealthy weight gain ($\geq 0.4$ kg/wk) by race for a woman with pre-pregnancy body mass index of 24 kg/m$^2$ and setting all other covariates at their medians or modes. 95\% confidence intervals are included. }
\label{fig:prob-race}
\end{figure}

Figure \ref{fig:odds-ratios} shows estimated odds ratios from the CPM with generalized raking; the figure also includes IPW estimates for the CPM and the naive CPM ignoring the phase-2 data. A few observations are notable: First, although for many covariates, odds ratios are similar between naive estimates and GR estimates, there are some substantive differences. In particular, GR estimates suggest tobacco use during pregnancy is associated with less weight gain whereas the result is largely null for the naive analysis. Naive estimates suggest having private insurance is associated with greater weight gain during pregnancy, whereas the GR estimate suggests no strong association (although with a wider confidence interval). Depression appears to be associated with less weight gain in the naive analysis, whereas for the GR analysis, the association appears to be in the opposite direction, although with a much wider confidence interval that includes the null. Confidence intervals for GR estimates are generally much wider than naive estimates, which is expected. In contrast, point estimates for GR and IPW are generally fairly similar (with the exception of Hispanic ethnicity) and 95\% confidence intervals for GR estimators are generally narrower than those of IPW estimators (average relative efficiency [RE], defined as the ratio of standard errors, of 0.67, range 0.17 [BMI coefficient] to 1.09). The exceptions are length of pregnancy (RE=1.09) and depression status (RE=1.06). Length of pregnancy was not in the phase-1 data and thus the estimated phase-1 influence function for this regression coefficient was not available for calibration -- hence, the absence of efficiency gains for this variable is understandable. The slight increase in the confidence interval width for the raked estimator for depression status could be because of the low PPV (0.32) for the phase-1 measure of depression, suggesting a low correlation between the naive influence function and the true influence function.

Women with greater BMIs tended to gain less weight during pregnancy. For two similar women, one with a BMI 5 kg/m$^2$ greater than the other prior to pregnancy, our GR estimator suggests that the relative odds of the woman with the higher BMI gaining more weight during pregnancy are 0.76 (95\% CI 0.70-0.82). This observation is consistent with general recommendations. %that heavier women gained less during pregnancy is consistent with recommendations by the United States Centers for Disease Control. %Weight gain recommendations during pregnancy are approximately 11.3 to 15.9 kg (corresponding with 0.29 to 0.41 kg/wk) for women of normal weight (BMI of 18.5 to 24.9 kg/m$^2$), but more for women who are underweight and less for women who are overweight. Our data suggests that in practice, holding all other factors constant, heavier women gained less during pregnancy, consistent with guidelines. 
We also investigated a potential non-linear relationship between BMI and weight change, by expanding BMI using restricted cubic splines with 3 knots. Such analyses are simple to perform using the CPM with GR. There was evidence of a non-linear relationship (Wald p-value=0.005). Figure \ref{fig:med-BMI} shows the estimated median weight gain as a function of BMI from the GR CPM, holding all other covariates at their medians or modes. The estimated median weight gain was fairly constant for BMIs $\leq 30$ kg/m$^2$, but then dropped for women $>30$ kg/m$^2$. There was also evidence of a non-linear relationship with length of pregnancy (p=0.030), so this variable was also expanded using restricted cubic splines with 3 knots (Figure \ref{fig:med-EGA}). For the other continuous risk factor, age, there was no evidence of a non-linear relationship (p=0.89). From the adjusted model with non-linear BMI and non-linear pregnancy length, the odds of gaining more weight during pregnancy decreased 27\% for a 10-year increase in age (odds ratio 0.73, 95\% CI 0.61, 0.88). 

Figure \ref{fig:prob-race} shows the probability of unhealthy weight gain, defined as $\geq 0.4$ kg/wk, by race for a woman with a pre-pregnancy BMI of 24 kg/m$^2$ (normal weight), and holding all other covariates at their medians or modes. Probabilities like this are easily extracted from the weighted CPM. These estimates suggest that approximately 23\% of White women with a normal pre-pregnancy BMI experienced unhealthy weight gain during pregancy. This estimated probability is much lower for Asian women (13\%) and lower for Black women (16\%).

The weighted likelihood for the CPM fit with the logit link ($-4349$) was slightly better than  those fit with the cloglog ($-4350$) and probit links ($-4355$) and much better than that fit with the loglog link ($-4380$). Results were very similar based on the weighted CPM with the probit link function and with the cloglog link function.  Results were also very similar when confidence intervals were computed using standard errors estimated from a bootstrap with 1000 replications that sampled with replacement from those sampled ($R=1)$ and those not sampled ($R=0$) within strata.

\section{Simulations}
\label{sec:sim}
Previous studies have investigated the performance of CPMs and compared them to  traditional modeling approaches \cite{liu2017, tian2020, li2023}. In brief, properly specified CPMs tend to have good finite sample performance with moderate (e.g., $n \approx 100$) or larger sample sizes. Although performance can be poor with severe link function misspecification, they appear to be fairly robust to moderate misspecification. Here, we present a limited set of simulations focusing on weighted estimation with CPMs with error-prone covariates. 

We generate two covariates, $X_1 \sim \mathcal{N}(0, 1)$ and $X_2 \sim \text{Bernoulli}(0.35)$, and define a latent outcome $Y^L = 0.1 + X_1\beta_1 + X_2\beta_2 + \epsilon$, where $\epsilon \sim \mathcal{N}(0, 1)$. The observed outcome is obtained via the transformation $Y = H(Y^L)$, where $H(y) = \mbox{Inv-}\chi^2(\Phi(y), 5)$. Here, $\Phi(\cdot)$ denotes the CDF of the standard normal distribution and $\mbox{Inv-}\chi^2(\cdot, 5)$ is the inverse of the CDF for a $\chi^2$-distribution with 5 degrees of freedom. This transformation, previously used in Tian et al. (2020), induces a right-skewed outcome with no obvious closed-form transformation function.  

We generate $N = 3,000$ observations for $X_1$, $X_2$, and $Y$, setting $\beta_1 = 0.5$ and $\beta_2 = -0.5$. To mirror what we observe in real data, we also simulate error-prone covariates, $X_1^*$ and $X_2^*$, and an outcome $Y^*$. To simulate our error-prone outcome, we first define our error-prone latent outcome $Y^{*L} = Y^L + \delta_Y$, where $\delta_Y \sim \mathcal{N}(0, \sigma^2_Y)$ and $\sigma^2_Y$ was set so that $Cor(Y^L, Y^{*L}) = 0.9$.  We then define $Y^* = H(Y^{*L})$ so that both the true and error-prone latent outcome have the same transformation.

Similar to how we introduce measurement error in $Y$, we simulate an error-prone $X_1^* = X_1 + \delta_X$ where $\delta_X \sim \mathcal{N}(0, \sigma^2_X) $ such that $Cor(X_1, X_1^*) \approx 0.9$.  To introduce measurement error in $X_2$, we simulate $X_2^*$ with sensitivity $\Pr(X_2^* = 1|X_2 = 1) = 0.90$ and specificity $\Pr(X_2^* = 0 | X_2 = 0) = 0.85$. Thus $(X_1^*, X_2^*, Y^*)$ are observed for the $N = 3,000$ individuals in our Phase 1 sample. A Phase 2 subcohort of size $n = 600$ is then selected using a stratified sampling design with 6 strata each of size $n_s = 100$ based on the percentiles of $Y^*$ (0-25, 25-75, 75-100) and levels of $X_2^*$ (0, 1). For the individuals in the Phase 2 subcohort, we assume that both the true ($X_1, X_2, Y$) and error-prone ($X_1^*,X_2^*, Y^*$)  measurements are observed.  For everyone not selected as part of the Phase 2 subcohort, ($X_1, X_2, Y$) is considered missing. This simulation setup is our \textit{base case} scenario. 

For estimating $\beta_1$ and $\beta_2$, we compare the following methods:
\begin{enumerate}
    \item[M1:] (True CPM) CPM using the true values $(X_1,X_2,Y)$ for all $N = 3,000$ observations.
    \item[M2:] (True LR) Linear regression on the correctly transformed outcome $Y^L = H^{-1}(Y)$ using the true values $(X_1,X_2,Y)$ for all $N = 3,000$ observations.
    \item[M3:] (P1 Only) CPM using the error-prone phase 1 data ($X_1^*, X_2^*, Y^*$) for all $N = 3,000$ observations.
    \item[M4:] (P2 Only) CPM using the true values $(X_1, X_2, Y)$ for the $n = 600$ individuals sampled in phase 2.
    \item[M5:] (IPW CPM) CPM using IPW for stratified sampling.
    \item[M6:] (GR CPM) CPM using the calibrated GR weights based on the influence functions from the P1 Only model.
    \item[M7:] (GR LR (log)) Linear regression on a log-transformed outcome using calibrated GR weights.
    \item[M8:] (GR LR (C)) Linear regression on the correctly transformed outcome $Y^L = H^{-1}(Y)$ using calibrated GR weights. 
\end{enumerate}

For the CPM models (M1, M3 - M5) we specify a probit link function for $G$. Models M1 and M2 assume that the true values, $(X_1, X_2,Y)$, are known for all individuals in Phase 1. M1 can be viewed as a ``true" CPM using phase 1 data; whereas M2 can be viewed as a ``gold standard" model. The calibrated GR weights for models M7 and M8 are constructed using the influence functions based on linear regression models using the error-prone phase 1 data and their respective transformation of the outcome. We chose a log transformation for M7 since our observed outcome $Y$ is right skewed. M8 is a correctly-specified two-phase model using the proper transformation, $H^{-1}(\cdot)$. In practice, M2 and M8 are unrealistic as such a transformation is typically not known \textit{a priori}. 

\begin{figure}
    \centering
    \includegraphics[scale = 0.2]{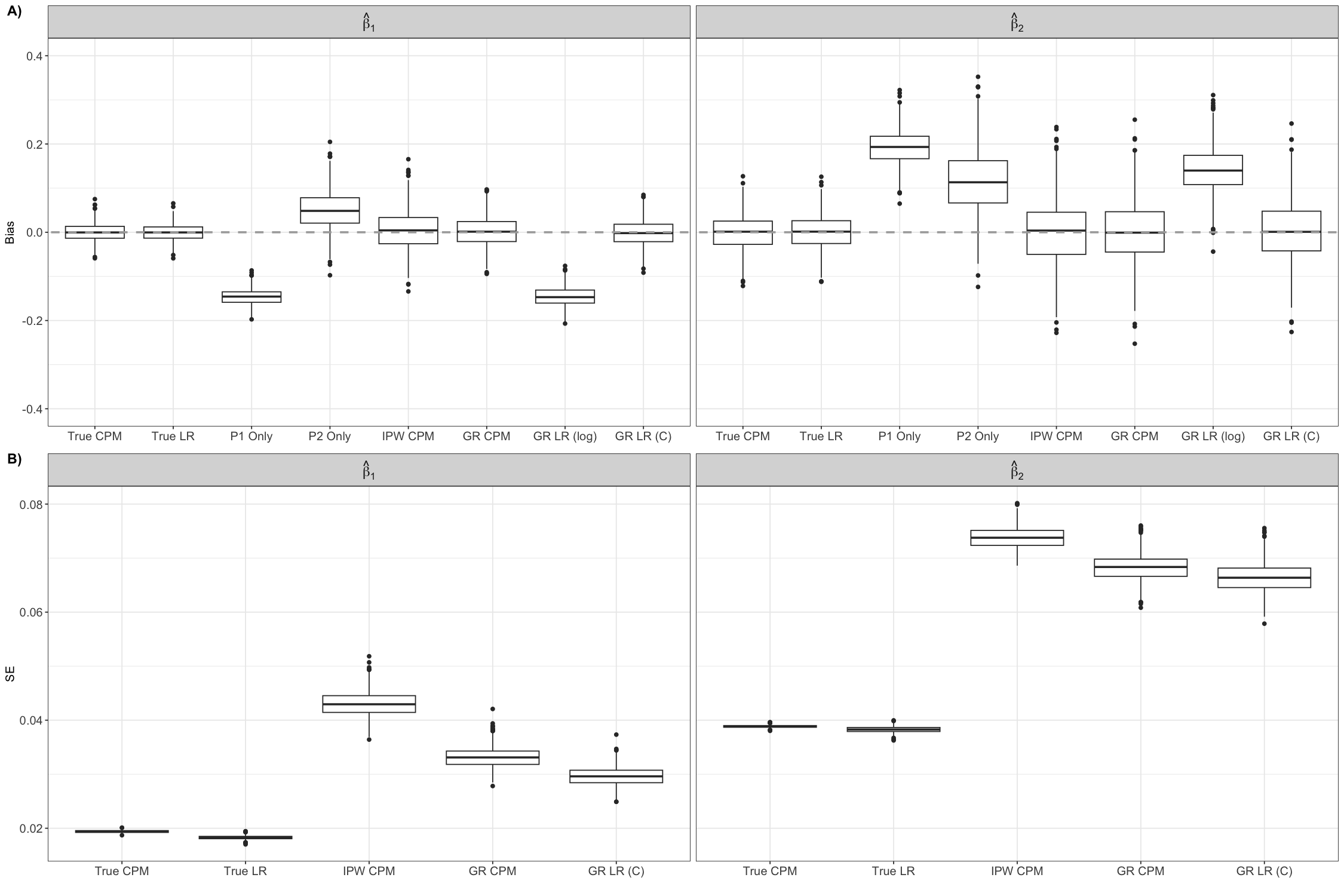}
    \caption{Simulation results under various estimation procedures. Estimation bias (Panel A) and estimates of the standard error (Panel B) are based on 1000 replicates. True: CPM or linear regression with properly specified transformation (LR) with $(Y,X_1,X_2)$ for everyone in phase 1; P1 Only: CPM with $(Y^*,X_1^*,X_2^*)$ for everyone in phase 1; P2 Only: CPM with $(Y,X_1,X_2)$ for everyone in phase 2; IPW CPM: CPM with IPW; GR CPM: CPM with GR weights; GR LR (log): Linear regression with log transformation and GR weights; GR LR (C): Linear regression estimator with properly specified transformation and GR weights. Simulation parameters are based on the \textit{base case} scenario as described in Section \ref{sec:sim}.}
    \label{fig:sim1}
\end{figure}

Figure \ref{fig:sim1} (top row) compares the bias for $\beta_1$ and $\beta_2$, i.e., $\hat{\beta}_j - \beta_j$, from the eight methods across 1000 replicates. The true CPM and linear regression models (M1 and M2), weighted versions of the CPM (M5 and M6), and the properly-specified linear regression using calibrated GR weights (M8) estimate the regression coefficients unbiasedly. Bias is observed in the other three models since 1) P1 Only (M3) uses the error-prone data, 2) P2 Only (M4) is an unweighted complete-cases analysis that does not take into account the biased phase-2 sampling based on $Y^*$, and 3) M7 misspecifies the transformation for linear regression. We also compare analytic standard error estimates among the five methods that unbiasedly estimate our regression coefficients (Figure \ref{fig:sim1} bottom row). As expected, our two gold standard models (M1 and M2) perform the best, although M2 is slightly more efficient because the correct transformation is assumed rather than estimated (M1).  Our proposed two-phase CPM using the calibrated GR weights (M6), on average, has smaller estimated standard errors when compared to the two-phase CPM using IPW (M5), underscoring the possible gain in efficiency via calibration. Lastly, while the properly specified linear regression model using calibrated GR weights (M8) performs better than the CPM using calibrated GR weights (M6), we reiterate that M8, as well as M2, assume that the \textit{correct }transformation is known ahead of time.

In addition, we examine the performance of CPM GR and IPW estimators of conditional quantiles and conditional exceedance probabilities. %Figure \ref{fig:sim3} shows the estimated median, 80th percentile, and exceedance probabilities, $Pr(Y > y)$, for $y=3$ and 6 conditional on $X_1=-1$ and $X_2=1$. The figure shows these estimates across 1000 simulations based on M5 (IPW CPM) and M6 (GR CPM). All estimators are consistent for the truth. 
We report estimation bias, empirical standard deviation and coverage probabilities for the estimated median, 80th percentile, and exceedance probabilities, i.e., $\Pr(Y > y)$, for $y=3$ and $6$ conditional on $X_1=-1$ and $X_2=1$ in Table \ref{tab:cond_sim}.  For completeness, we also included these metrics for estimating $\betabf = (\beta_1, \beta_2$).  Both methods provide unbiased estimates for the various quantities.
The GR estimator is more efficient than the IPW estimator, demonstrating the benefits of calibrating weights using the naive influence functions for $\betabf$. Notably, the efficiency gains are greater for $\betabf$ than for the quantiles and exceedance probabilities, as these latter quantities are functions of both ($\alphabf,\betabf$).  Coverage is close to the nominal level for all estimators, suggesting that the proposed analytic variance estimators are well behaved.

\begin{figure}
    \centering
    \includegraphics[scale = 0.15]{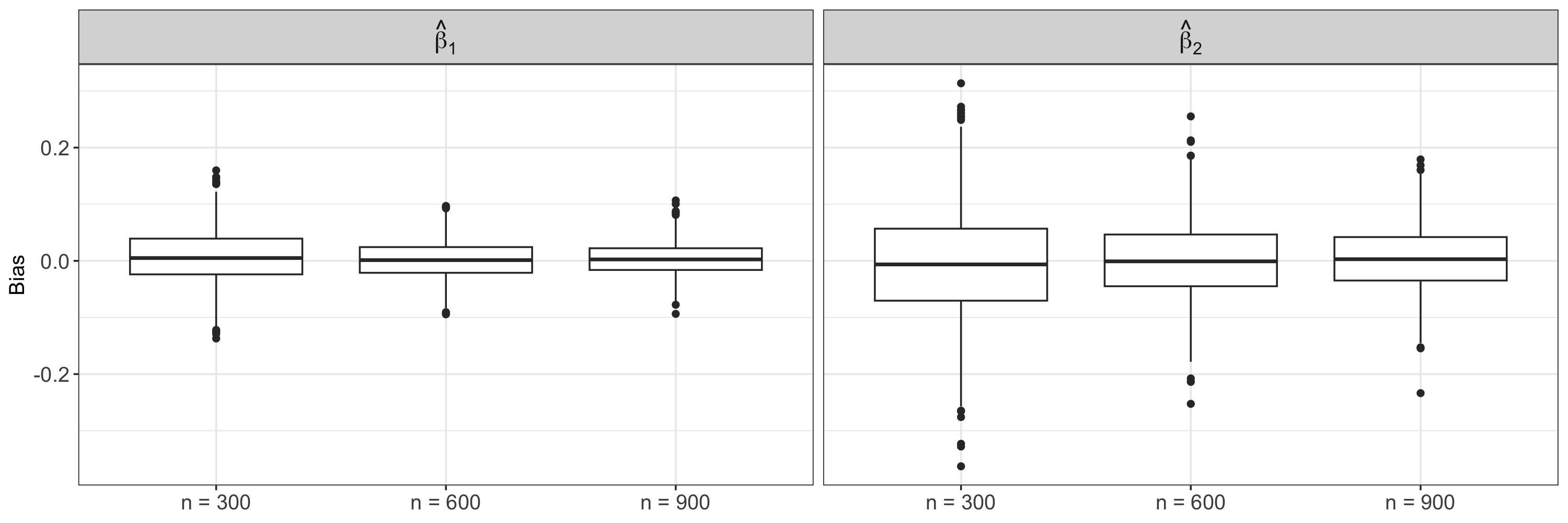}
    \caption{ Simulation results comparing CPM with generalized raking weights under various phase-2 sample sizes ($n = 300, 600, 900$). Boxplots of bias ($\hat{\beta}_j - \beta_j$, $j = 1, 2$) are based on 1000 simulation replicates. }
    \label{fig:sim2}
\end{figure}

Lastly, we compare the performance of generalized raking across different phase 2 sample sizes and across different levels of measurement error in our covariates. Figure \ref{fig:sim2} shows boxplots of bias in estimating $\beta_1$ and $\beta_2$ as a function of the phase-2 sample size ($n = 300, 600, 900$) while keeping all other settings as in our base case scenario. As expected, we gain precision in our estimates as the phase-2 sample size increases. We report the estimated standard deviations, averaged over 1000 replicates, across different levels of measurement error in $X_1$ and $X_2$ in Table \ref{tab:me_sim}. Both IPW and GR behave similarly when the phase-1 data are uncorrelated with our phase-2 data ($Cor(X_1, X_1^*) = 0$, Sensitivity = Specificity = 0.5). As correlation increases, GR is more efficient in estimating $\beta_1$ and $\beta_2$. Thus our simulations illustrate that efficiency improves with better (i.e., more highly correlated) auxiliary variables, but that calibration with poor auxiliary variables does not perform worse than IPW.

%Our current simulation assumes that the latent outcome was measured with error but that both the true and error-prone latent outcome have the same transformation. We performed additional simulations where the transformation itself was measure with error. 

\begin{comment}
\begin{figure}
    \centering
    \includegraphics[scale=0.15]{plots/sim3.png}
    \caption{[I THINK YOU CAN REMOVE THIS FIGURE AND JUST ADD THE MEAN BIAS TO TABLE 2.] Estimated quantiles (Panels A and B) and exceedance probabilities (Panels C and D) averaged over 1,000 simulations. True quantiles and exceedance probabilities were estimated empirically based on 1,000,000 realizations and illustrated on the figure as the grey dotted line. Coverage probabilities are in parentheses. Simulation parameters are based on the \textit{base case} scenario as described in Section \ref{sec:sim}.} 
    \label{fig:sim3}
\end{figure}
\end{comment}

\begin{table}
\caption{Bias, empirical standard deviation (Std. Dev.), and coverage probabilities (Coverage) for 95\% confidence intervals based on estimated standard errors for regression coefficients, conditional quantiles, and  exceedance probabilities averaged over 1000 replicates. IPW: CPM using the IPW for stratified sampling; GR: CPM using the calibrated GR weights. Quantiles and exceedance probabilities were estimated conditional on $X_1 = -1$ and $X_2 = 1$. True quantiles and exceedance probabilities were estimated empirically based on 1,000,000 realizations. Simulation parameters are based on the \textit{base case} scenario as described in Section \ref{sec:sim}.}
    \label{tab:cond_sim}
    \centering
    \begin{tabular}{rcccccc}
       & \multicolumn{2}{c}{Bias} &  \multicolumn{2}{c}{Std. Dev.}&\multicolumn{2}{c}{Coverage}\\
 & IPW& GR& IPW& GR& IPW&GR\\
 \hline
 $\beta_1$& 0.004& 0.002& 0.044& 0.033& 0.959&0.954\\
 $\beta_2$& <0.001& <0.001&  0.073& 0.069& 0.946&0.941\\
 Median& 0.003& 0.005&    0.134& 0.128&0.954& 0.950\\
 80th Percentile& -0.004& <0.001&    0.241& 0.227&0.939& 0.942\\
 $\Pr(Y>3)$& <0.001& <0.001&    0.029& 0.028&0.952& 0.949\\ 
 $\Pr(Y >6)$& <0.001& <0.001&  0.013& 0.012& 0.952&0.944\\ 
 \hline
    \end{tabular}
\end{table}

\begin{table}
    \caption{Empirical standard deviation averaged over 1000 replicates across different levels of measurement error in $X_1$ and $X_2$. IPW: CPM using the IPW for stratified sampling; GR: CPM using the calibrated GR weights. $\rho_x = Cor(X_1, X_1^*)$; Se = Sensitivity = $\Pr(X_2^* = 1|X_2 = 1)$; Sp = Specificity = $\Pr(X_2^* = 0 | X_2 = 0)$. All other simulation parameters (e.g., $N, n, \beta_1, \beta_2$) are the same as in the base case scenario.}
    \centering
    \begin{tabular}{rcccc}
         &  \multicolumn{2}{c}{$\beta_1$}&\multicolumn{2}{c}{$\beta_2$}\\
 & IPW& GR& IPW&GR\\
 \hline
         $\rho_x = 0$, Se = 0.5, Sp = 0.5&   0.043& 0.043&0.085& 0.086\\
         $\rho_x = 0.8$, Se = 0.8, Sp = 0.75&   0.046& 0.042&0.082& 0.082\\
         $\rho_x = 0.9$, Se = 0.9, Sp = 0.85&   0.044& 0.033&0.073& 0.069\\ 
         \hline
    \end{tabular}
    \label{tab:me_sim}
\end{table}

\section{Discussion}
\label{sec:discussion}
In this manuscript, we have combined two novel analysis methods, the CPM and generalized raking, to obtain robust and efficient estimates of association in a two-phase sampling setting to address measurement error across multiple variables. These methods were motivated by a study using error-prone EHR data together with validated data on a subsample of records. Methods of this type are important as routinely collected data of lower quality continue to be used for biomedical research.

Generalized raking estimators can substantially improve efficiency over inverse probability weighted estimators. This was seen in both our simulations and in our application study. For example, the standard error for the GR estimator of the BMI log-odds ratio was 0.17 times the standard error for the IPW estimator. This remarkable gain in efficiency comes from incorporating information from the larger phase-1 cohort into the estimation. Even though EHR variables are measured with errors, substantial information still exists in these error-prone measures, particularly if the correlation between the error-prone and corrected measures is high. It is important to note that the efficiency gains made by GR estimators in a sense come for free: no additional assumptions are made over those made by IPW, and GR estimators are guaranteed, at least asymptotically, not to be less efficient than their IPW counterparts. %Our current approach does not calibrate using the potentially high-dimensional naive influence function for $\alphabf$ and acknowledge that GR estimators can suffer from instability if number of raking variables becomes too large relative to sample size.

The CPM is also an important method that is worth having in one's analytical toolbox. Long used for the analysis of ordinal data, its use with continuous and mixed data has recently been recognized. Critically, statistical software advances over the last ten years have permitted fitting these models to data with hundreds of thousands of unique continuous outcomes. Our developments to incorporate inverse probability weights and in particular, generalized raking, into CPMs represent important extensions that permit the use of these robust models to settings with two-phase sampling or survey sampling. 

Our study focused on settings with continuous outcomes and errors in multiple variables, because that was the nature of our data example. However, the methods developed here could be applied in straightforward manners to other settings such as those with errors in fewer variables, those with a partially missing expensive predictor, and those with ordinal or mixed type outcomes. The setting with a partially missing expensive predictor often arises because researchers only have funds to measure a covariate on a subsample of records. In this setting, there is no surrogate measure for $X$ (e.g., $X^*$), so raking could be performed using influence functions based on predicted / imputed estimates of $X$ (e.g., $\hat X$, perhaps based on multiple imputations). In settings with ordinal outcomes, the analyst could likely calibrate weights using not only the influence functions for the $\betabf$ coefficients, but also for the $\alphabf$ coefficients. We were unable to do this in our setting with a continuous response variable because of the high dimension of $\alphabf$; GR estimators can suffer from instability if the number of raking variables becomes too large relative to the sample size. In the lower dimension setting with ordinal $Y$, calibrating on the influence functions for $\alphabf$ in addition to those for $\betabf$ is likely to improve efficiency -- particularly for estimands that are functions of $\alphabf$ such as exceedance probabilities. Consequently, despite the gains in efficiency that we saw using GR over IPW, there may be additional potential efficiency gains to be made with continuous outcomes. For example, perhaps one could calibrate on a subset of the influence functions for $\alphabf$; or one could categorize $Y^*$ (e.g., using deciles), fit a CPM in the phase-1 data to this lower dimensional ordinal response, calibrate weights using naive influence functions for $\alphabf$ from this lower dimensional model, and then apply these calibrated weights to fit the CPM to the phase-2 data on the original continuous scale. We investigated this strategy in a small set of simulations and found very minor efficiency gains for quantile estimators and exceedance probabilities (see Supplemental Table \ref{tab:supp_sim}). 

%A study of these and other strategies to potentially improve efficiency of CPM with GR represents an interesting direction for future research. % --  over calibrating only on the influence function for $\betabf$.

%Our current approach does not calibrate using the potentially high-dimensional naive influence function for $\alphabf$ and acknowledge that GR estimators can suffer from instability if number of raking variables becomes too large relative to sample size.

\bibliographystyle{abbrv}
\bibliography{references}

\section*{Appendix}
\renewcommand{\thefigure}{A\arabic{figure}}
\setcounter{figure}{0}

\renewcommand{\thetable}{A\arabic{table}}
\setcounter{table}{0}

\subsection*{Efficient calculation of influence functions}
Without loss of generality, we will focus on calculating the weighted influence function using phase-2 data. We have
\begin{align}
    \label{eq:influence}
    \hbf_i(\thetabf) = -\hat{\mathcal{I}}_w^{-1}U_i(\thetabf),
\end{align}
where $\hat{\mathcal{I}}_w$ is the weighted estimate of the population Fisher information and $U_i(\thetabf) = \frac{\partial}{\partial \thetabf} l_i(\thetabf)$ is the score component of the $i$th individual. Assuming no ties in the observed outcomes, $\hat{\mathcal{I}}_w$ is a $(n + p - 1) \times (n + p - 1)$ matrix and $U_i(\thetabf)$ is $(n + p - 1)$-dimension vector. We can further partition $\hat{\mathcal{I}}_w$:

\begin{equation}
\hat{\mathcal{I}}_w = 
\begin{pmatrix}
\begin{matrix} P & W \\ W^T & Q \end{matrix} 
\end{pmatrix},
\end{equation}
where $P$ is a $(n - 1) \times (n - 1)$ tri-diagonal matrix, $W$ is a $(n - 1) \times p$ dense matrix and $Q$ is a $p \times p$ dense matrix \cite{liu2017}. Also, $U_i(\thetabf) = (U_{\alpha i}(\thetabf), U_{\beta i}(\thetabf))$. 

Assume, for now, that we are only interested in computing the influence functions for $\betabf$, $h_{\beta i}(\thetabf)$. One can show that, 
\begin{align}
    h_{\beta i}(\thetabf) = (Q^{-1} - W^TP^{-1}W)^{-1}(U_{\beta i}(\thetabf) - W^TP^{-1}U_{\alpha i}(\thetabf)).
\end{align}
Although $P$ is $(n - 1) \times (n - 1)$, solving the tridiagonal system $PX=W$ is $\mathcal{O}(n)$ for fixed $p$.

If one wants to perform inference on quantities based on the outcome conditional distribution (e.g., the conditional mean and quantiles), one must also get a standard error estimate for $\hat{\alphabf}$, which requires the influence functions for $\alphabf$

\begin{align*}
      h_{\alpha i}(\thetabf) = P^{-1}(U_{\beta i}(\thetabf) - W h_{\beta i}(\thetabf)).
\end{align*}

Computing $h_{\alpha i}(\thetabf)$ is $\mathcal{O}(n^2)$ and requires storing a dense $(n-1) \times n$ matrix.

\newpage
\subsection*{Additional tables and figures}

\begin{figure}[h]
\includegraphics[scale=1]{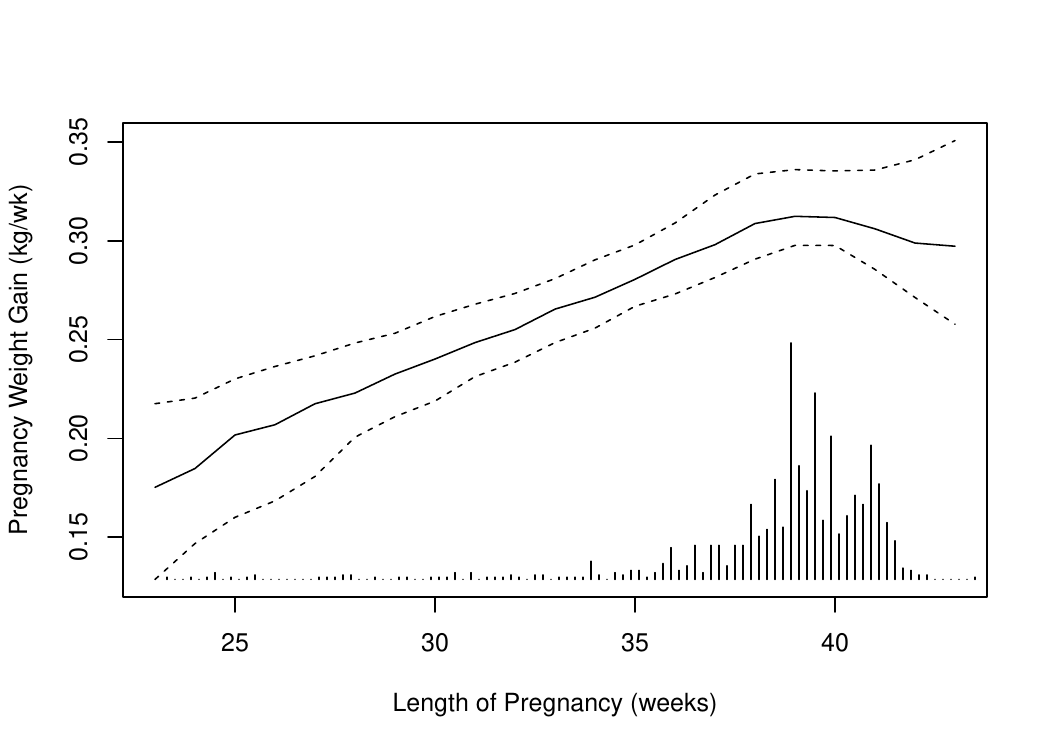}
\caption{Estimated median weight change (kg/wk) during pregnancy as a function of length of pregnancy. 95\% confidence intervals for the median are included. A histogram of pregnancy lengths in the unweighted phase-2 cohort is shown at the bottom of the figure. All other covariates are set at their medians or modes.}
\label{fig:med-EGA}
\end{figure}

\begin{table}[h!]
\caption{Results from simulations investigating the impact of calibrating weights using naive influence functions for $\bm \alpha$ from models with categorized $Y^*$. Simulation parameters are those of the \textit{base case} scenario as described in Section \ref{sec:sim}. GR ($k$) denotes generalized raking estimators with weights calibrated on phase-1 influence functions for $\bm \beta$ from a CPM of $Y^*$ on $(X_1^*,X_2^*)$ and phase-1 influence functions for $\bm \alpha$ from a CPM of binned $Y^*$ on $(X_1^*,X_2^*)$ with $k$ equally sized bins, for $k=$ 5, 10, and 20. GR(0) denotes no calibration based on $\bm \alpha$ (i.e., based on $\bm \beta$ only). Quantiles and exceedance probabilities were estimated conditional on $X_1 = -1$ and $X_2 = 1$. True quantiles and exceedance probabilities were estimated empirically based on 1,000,000 realizations. Empirical standard deviation (Std. Dev.) and coverage probabilities (Coverage) for 95\% confidence intervals for regression coefficients, conditional quantiles, and  exceedance probabilities averaged over 1000 replicates are reported.}
    \label{tab:supp_sim}
    \centering
    \begin{tabular}{rcccccccc}
        &  \multicolumn{4}{c}{Std. Dev.} & \multicolumn{4}{c}{Coverage}\\
 & GR (0) & GR (5) & GR (10) & GR (20)  & GR (0) & GR (5) & GR (10) & GR (20) \\
 \hline
 $\beta_1$& 0.033 & 0.035 & 0.034 & 0.033 & 0.954 & 0.942 & 0.947 & 0.947 \\
 $\beta_2$&  0.069 & 0.070 & 0.070 & 0.070 & 0.941 & 0.944 & 0.942 & 0.938 \\
 Median&    0.128 & 0.126 & 0.125 &  0.125  & 0.950 & 0.946 & 0.938 & 0.935 \\
 80th Percentile &   0.227 & 0.216 & 0.214 & 0.213 & 0.942 & 0.944 & 0.938 & 0.933  \\
 $\Pr(Y>3)$&   0.028 & 0.027 & 0.027 & 0.027 & 0.949 & 0.945 & 0.948 & 0.943 \\ 
 $\Pr(Y >6)$&  0.012 &  0.012 & 0.012 & 0.012 & 0.944 & 0.933 & 0.945 & 0.947 \\ 
 \hline
    \end{tabular}
\end{table}

\subsection*{R code for simulations}
R code for scripts to reproduce results and figures in Sections 4 and 5 can be found at https://github.com/erickawaguchi/two-phase-cpm.
\end{document}